\documentclass{article}
\usepackage[a4paper, margin=1in]{geometry}

\usepackage{graphicx,psfrag,epsf}
\usepackage{url} 
\usepackage{adjustbox}
\usepackage{multirow}
\usepackage{hhline}

\usepackage{bm}
\usepackage{mathtools}
\usepackage{comment}

\RequirePackage{amsthm,amsmath,amsfonts,amssymb}
\RequirePackage[authoryear]{natbib}
\RequirePackage[colorlinks,citecolor=blue,urlcolor=blue]{hyperref}
\RequirePackage{graphicx}

\newcommand*\mean[1]{\bar{#1}}
\renewcommand{\vec}{\bm}

\newcommand{\T}{\mathcal{T}}
\newcommand{\SSS}{\mathcal{S}}
\newcommand{\LL}{\mathcal{L}}

\newcommand{\M}{\mathcal{M}}

\begin{document}

\providecommand{\keywords}[1]
{
  \small	
  \textbf{\textit{Keywords---}} #1
}

\title{Flexible Instrumental Variable Models With Bayesian Additive Regression Trees}

\author{\\[2ex]Charles Spanbauer\footnote{Corresponding author, spanb008@umn.edu}~ and Wei Pan\\[4ex] \large Division of Biostatistics, University of Minnesota, Minneapolis, MN}

\date{}

\maketitle

\begin{abstract}
Methods utilizing instrumental variables have been a fundamental statistical approach to estimation in the presence of unmeasured confounding, usually occurring in non-randomized observational data common to fields such as economics and public health. However, such methods usually make constricting linearity and additivity assumptions that are inapplicable to the complex modeling challenges of today. The growing body of observational data being collected will necessitate flexible regression modeling while also being able to control for confounding using instrumental variables. Therefore, this article presents a nonlinear instrumental variable regression model based on Bayesian regression tree ensembles to estimate such relationships, including interactions, in the presence of confounding. One exciting application of this method is to use genetic variants as instruments, known as Mendelian randomization. Body mass index is one factor that is hypothesized to have a nonlinear relationship with cardiovascular risk factors such as blood pressure while interacting with age. Heterogeneity in patient characteristics such as age could be clinically interesting from a precision medicine perspective where individualized treatment is emphasized. We present our flexible Bayesian instrumental variable regression tree method with an example from the UK Biobank where body mass index is related to blood pressure using genetic variants as the instruments.
\end{abstract}


\section{Introduction}
\label{sec:intro}

The presence of unmeasured confounding, particularly in non-randomized observational data, can bias the estimated effect of an exposure on an outcome and make the interpretation of scientific results difficult. Obtaining unbiased results in such a case has traditionally been done through instrumental variable (IV) methods. IV methods incorporate an additional variable, the instrument, to induce exposure variability that is independent of the confounders thereby yielding unbiased estimation of the exposure effect \citep{StocTreb03}. However, this strategy is only valid if the instrument satisfies three properties, usually known as the instrumental variable assumptions. Fields such as economics and public health collect vast quantities of observational data with confounding issues and so researchers in these fields frequently turn to IV analyses to obtain unbiased results.

The three properties central to obtaining unbiased inference are usually known as the instrumental variable assumptions. First, the instruments must be predictive of the exposure in order to induce exposure variability. Second, the instruments must be uncorrelated with the confounders which ensures that this induced variability is independent of the confounders. Finally, there must be no direct effect between the instruments and the outcome, i.e. the instruments only affect the outcome through the exposure. Taken together, these properties imply that associating the instrument-induced exposure to the outcome will be free of confounder influence, yielding an unbiased estimate of the exposure effect that can be interpreted causally. Unfortunately, only the first of these assumptions is testable with the gathered data. As such, it is a notoriously difficult and challenging problem to find valid instruments for a particular research question and it usually requires domain knowledge about the data generating mechanism. For example, a classical question in economics seeks to correctly identify the extent to which education levels correlate with wages, but that relationship is hypothesized to be confounded by the innate ability levels of the subjects \citep{Card99}. Not only is ability level unmeasured, but there may not even be consensus as to what constitutes innate ability. Nevertheless, IV methods can be used to assess this research question because ability level only needs to be defined enough to justify the choice of instrument and whether it satisfies the IV assumptions. The geographic proximity to a two or four year college has been used as an instrument in this case, the choice of which is based on domain knowledge within the economics and education literature.

In biomedical research, IV methods have played a large and increasingly important role in the analysis of observational data which is plagued by confounding. Furthermore, the volume and complexity of these datasets are growing over time, necessitating the development of more sophisticated IV methods. See \cite{BaioChen14} for an overview of IV methods in biomedical sciences. For example, the use of IV methods has recently shown promise in the area of statistical genetics where a causal genetic relationship can be established by using genetic variants as instruments, specifically single-nucleotide polymorphisms (SNPs), for various traits or phenotypes. This is commonly called Mendelian randomization (MR) because genetic instruments are used as a proxy for the true randomization that is absent in observational data. The IV assumptions in such a case are biologically plausible as long as these SNPs are carefully chosen so as to not be associated with environmental confounders.

Most IV approaches are based on the linear two-stage least-squares (2SLS) methodology which predicts the exposure from the instruments in a first stage regression and then uses the predicted exposure to estimate the unbiased effect of the exposure on the outcome in a second stage regression. Based on the IV assumptions, the variability in the predicted exposure should not be associated with the confounders and so the estimated effect of the exposure will be unbiased. Therefore, the results from this second stage regression will have a causal interpretation assuming the validity of the instruments. However, all 2SLS methods rely on linearity assumptions that can be problematic \citep{Horo11}. This is particularly true in MR methods that use genetic variants as instruments. Work by \cite{GrinWall21} suggests that linear models may not be sufficient to capture the effect of genetic variants on different traits. Relaxing the linearity assumption could lead to improved prediction in the first stage and improved inference via better power in the second stage. Therefore, methods that relax linearity and additivity assumptions may prove beneficial over ones that do not in the statistical genetics context.

Beyond genetic variants, the traits of interest themselves may have relationships hypothesized to be nonlinear, the estimates of which can provide a more complete picture of the underlying regression relationship. There are a variety of examples in biomedical research \citep{PeteMaye15,ScarAndr17}, including include body mass index with blood pressure, an example that is explored in Section~\ref{sec:ukbb} using the UK Biobank dataset. There are additional examples where nonlinear effects have been investigated in fields including, but not limited to: economics \citep{BurkHsia15,BoteEger19}, marketing \citep{TuuHo10,YinBond17}, and political science \citep{Kiel00,LipsPadi21}.

Homogeneity of the exposure effect has been called an implicit assumption of IV methods \citep{Lous18}, an assumption increasingly regarded as untenable in clinical settings whose focus is on precision medicine and individualized treatment. Individualizing the estimation of the exposure effect by allowing flexible interactions with patient characteristics could be incredibly useful in forming beneficial treatment strategies. Estimating the global effect of a trait without considering age may give an incomplete picture of the results. For example, age is known to have a heterogenous impact on genetic risk for various diseases \citep{JianHolm21}. The National Academy of Medicine lists ``increasing evidence generation'' as one of five main challenges facing researchers in the area of precision medicine \citep{DzauGins16}.

To this end, nonparametric IV regression, which relaxes the linearity and additivity assumptions of 2SLS, has been a popular research area of research for statisticians and econometricians. Examples of such methodologies include \cite{NewePowe03} which are based on series approximations, \cite{BurgDavi14} which stratifies the exposure and calculates the causal effect within each strata, and \cite{GuoSmal16} which adjusts for the first stage errors in the second stage model, something known as control function estimation. In another example, \citep{ChetWilh17} impose monotonicity which can help stabilize the high variance typically found with series approximations. Stratifying the exposure is an effective strategy but requires the cut points to be chosen by the analyst. This method also utilizes the ratio estimator which can only have a single instrument. Finally, these methods do not allow for general heterogeneity in the exposure effect unless the interactions are specified by the analyst.

Bayesian inference has also been used for IV models as an alternative to methods based on 2SLS or control functions. For examples, see \cite{LopePols14} and \cite{WiesHisg14}. One benefit is the inherent regularization that is provided by the prior which can help guard against the bias that arises when the instruments are not predictive of the exposure, a violation of the IV assumptions. Another benefit is that both stages are modeled simultaneously with a full probability distribution. This means that the uncertainty inherent in the first stage will be propagated to the second stage during estimation, in contrast to 2SLS where the uncertainty associated with the first stage predictions is not accounted for. Indeed, Bayesian inference allow for the uncertainty for \textit{any} quantity of interest to be estimated when MCMC is employed to sample from the posterior distribution. One final benefit is that Bayesian inference allows for easy extendibility through hierarchical modeling in order to handle more complicated analysis scenarios.

To allow for both nonlinearity and heterogeneity, Bayesian additive regression trees \citep{ChipGeor10}, or BART, priors can be incorporated to the Bayesian framework for IV analysis. BART has shown promise in flexibly estimating general regression relationships without needing any assumptions on the functional form such as higher-order or interaction terms. This article introduces such a heterogenous nonparametric model for use in instrumental variable analyses called npivBART-h, along with simple default settings for the prior so that the method can be used easily and efficiently with minimal learning curve. BART has experienced a usage increase over the past decade along with the development of multiple extensions to handle increasingly complex data. For example, BART has support for many alternative outcome types such as binary, count \citep{Murr21}, and survival \citep{SparLoga16} including competing risks \citep{SparLoga20} and recurrent events \citep{SparRein20}. Repeated measure outcomes can also be handled through a random effect specification \citep{TanFlan16,SpanSpar21} that is combined with a BART prior to estimate nonlinear fixed effects. There has also been research toward applying BART in high-dimensional sparse situations \citep{RockVand17,Line18}. Finally, work has been done on applying BART to precision medicine using individidualized treatment rules \citep{LogaSpar19}. All of these extensions could easily be incorporated into the npivBART-h framework, broadening the scope of its applicability for causal inference.

This article is organized as follows. In Section~\ref{sec:ivBART}, the npivBART-h model is specified with advice for setting the priors. Section~\ref{sec:sim} presents a simulation study showing the unbiased inference for these methods in the presence of confounding. Estimation consequences in the presence of weak instruments, an IV assumption violation, is explored through simulation as well. The results from the UK Biobank data relating body mass index to blood pressure with heterogeneity in age and sex using npivBART-h are presented in Section~\ref{sec:ukbb}. The paper concludes with a discussion in Section~\ref{sec:conc}.

\section{Instrumental Variable Analysis w/ BART}
\label{sec:ivBART}

The IV model using Bayesian Additive Regression Trees is described in this section. A brief overview of parametric Bayesian linear IV analysis methods is presented and then our extension to nonlinear relationships using BART is given in Section~\ref{ssec:2sls+npivBART}. The default regularization priors used in traditional BART are discussed as well as how these priors can be adapted to the IV setting in Section~\ref{ssec:prior}. Section~\ref{ssec:lessgeneral} gives a brief treatment of simpler semiparametric models that can estimate interpretable linear effects while Section~\ref{ssec:dpm} delves into the Dirichlet process mixture model specification of the errors.

\subsection{Linear Simultaneous Equations Model and npivBART-h}
\label{ssec:2sls+npivBART}

The instrumental variables strategy commonly used in Bayesian IV analyses defines a linear simultaneous equations model as
\begin{align}
  t_i&=\vec{\gamma}'\vec{z}_i+\epsilon_{t_i}\label{eqn:tsls1}\\
  y_i&=\beta t_i + \vec{\delta}'\vec{x}_i+\epsilon_{y_i},\label{eqn:tsls2}
\end{align}
where $\beta$ is the causal effect of interest assuming the IV assumptions are met. The dependency between equations are adjusted for through the error specification $\vec{\epsilon}_i=(\epsilon_{t_i},\epsilon_{y_i})'\sim\text{N}_2(\vec{0}_2,\Sigma)$. The covariance between the errors controls the degree of confounding. Equation~\eqref{eqn:tsls1} represents the first stage model relating instruments $\vec{z}_i$ with the exposure of interest $t_i$. Equation~\eqref{eqn:tsls2} represents the second stage model with the outcome $y_i$ being modeled as a linear function of the exposure and some other measured covariates $\vec{x}_i$. We assume centered outcomes and so the intercepts are omitted for simplicity. This model is defined in \cite{RossAlle12} and \cite{LopePols14}.

The goal is to relax the linearity assumption in the above model while still retaining unbiased estimation of the causal exposure effect. This can be done with:
\begin{align}
  t_i&=f_1(\vec{z}_i)+\epsilon_{t_i}\label{eqn:npivbart1}\\
  y_i&=f_2(t_i,\vec{x}_i)+\epsilon_{y_i}\label{eqn:npivbart2}
\end{align}
where $\vec{\epsilon}_i$ is defined as above. The model that Equations~\eqref{eqn:npivbart1} and \eqref{eqn:npivbart2} specify is analogous to the above model, but $f_2$ replaces $\beta$ as the causal exposure effect of interest. Note that the other covariates $\vec{x}_i$ are incorporated into function $f_2$ which yields a heterogenous causal effect that is not necessarily linear. BART is used to place such a nonlinear prior on $f_1$ and $f_2$, which implies the following specification:
\begin{align*}
  f_1(\vec{z}_i)&\approx\sum_{h=1}^{H_t}g(\vec{z}_i;\T_h,\M_h)\\
  f_2(t_i,\vec{x}_i)&\approx\sum_{h=1}^{H_y}g(t_i,\vec{x}_i;\SSS_h,\LL_h).
\end{align*}
Here, $g$ are recursively defined piecewise constant functions otherwise known as regression trees. In this definition, $\T_h$ and $\SSS_h$ represent the probabilistic structure for partitioning the predictor space in function $g$. $\M_h$ and $\LL_h$ represent the terminal node values that serve as the output of function $g$. For simplicity let $(\T,\M)$ denote the set of all $H_t$ trees approximating $f_1$ and let $(\SSS,\LL)$ be denoted similarly for $f_2$. In a Bayesian sense, specifying priors on $f_1$ and $f_2$ can be done by specifying priors on $(\T,\M)$ and $(\SSS,\LL)$ respectively.

\subsection{BART and npivBART-h Regularization Priors}
\label{ssec:prior}

All aspects of the BART prior involves some degree of regularization which is what helps guard against overfitting. In particular, \cite{ChipGeor10} provide default prior settings that are lightly driven by the data, but perform very well without any cross-validation, although using cross-validation can improve performance.

Let us first consider the traditional BART regression model where the target of inference is $E[y_i|\vec{x}_i]=f(\vec{x}_i)$ and the regression relationship $f$ is approximated by $H$ regression trees in the ensemble $(\T,\M)$. For the prior, the trees in this ensemble are taken to be independent. Then, for each $(T_h,\M_h)$ pair, the prior is broken down into components $\T_h$ and $\M_h|\T_h$. Partitioning the pair in this way is required because number of terminal nodes in $\M_h$ is not known until $\T_h$ is determined.

The prior on the individual regression trees $\T_h$ involves three components: the probability of a node being an interior node at depth $\delta$, the probability of a splitting rule at an interior node using predictor $j$, and the probability of the splitting rule at an interior node using cut point $c$ of predictor $j$. The first is set to decay as the trees grow deeper, favoring simpler trees which helps to avoid overfitting and provides regularization. The second component by default is set to be $p^{-1}$ for all predictors $j=1,\dots,p$. The third component is set by creating an equidistant grid with equal probability for each predictor. This can be extended to binary predictors without loss of generality and categorical / ordinal predictors with a sufficient transformation to dummy indicators. Full details are found in \cite{ChipGeor10}.

Given the tree structure, the terminal node values $\M_h|\T_h$ also need a prior. Let the elements of $\M_h$ be $\nu_{hb}$ for $b=1,\dots,B_h$ where $B_h$ is the number of terminal nodes in tree $h$. A normal prior centered around the sample mean (or $0$ if the outcome is centered). The variance of this prior is set so that a certain percentage of the prior captures the range of the outcomes $\max(y_i)-\min(y_i)$, reflecting the fact that predictions should not be made outside of this observed range with a high probability. This regularization is quantified by $k$ which controls the prior variability in the the end predictions. For example, $k=2$ implies that the range of the data covers the central $95\%$ of the prior while $k=3$ implies that this prior coverage is $99.9\%$.

While these prior settings use the data in the form of the sample mean and range, it should be noted that using the sample mean is regularizing the regression relationship to a flat line which represents the null model in a regression setting. Using the range ensures that the end predictions from our model are calibrated to reasonable values. In the npivBART-h setting, the first stage model can be scaled to the range of the exposure $\max(t_i)-\min(t_i)$ and, as in traditional BART, the outcome $\max(y_i)-\min(y_i)$. Other components of the BART prior can be modified as context dictates, but that will be unnecessary for the rest of this article.

\subsection{Semiparametric ivBART Models}
\label{ssec:lessgeneral}

Less flexible models are sometimes worth considering because of their added interpretability. For example, suppose that $f_2(t_i,\vec{x}_i)=\beta t_i+f_{22}(\vec{x}_i)$ or alternatively $f_2(t_i,\vec{x}_i)=f_{21}(\vec{x}_i)t_i+f_{22}(\vec{x}_i)$. The former (ivBART-g) estimates a global linear exposure effect after accounting for the nonlinear effect of the measured confounders. The latter (ivBART-h) estimates a linear exposure effect that can vary between observations based on their measured characteristics $\vec{x}_i$. In particular, the latter estimate could be useful in the biomedical sciences where precision medicine is a major focus of research. Setting the variance of the prior for $\beta$ in ivBART-g can be done by in the same manner as the previous subsection, but using $\{\max(y_i)-\min(y_i)\}/\{\max(t_i)-\min(t_i)\}$ instead of the range of the outcomes. This modification reflects the fact that $\beta$ represents a slope term. Method ivBART-g has been discussed in \cite{McCuSpar22}.

Finally, npivBART-h has a global counterpart (npivBART-g) that is specified as $f_{2}(t_i,\vec{x}_i)=f_{21}(t_i)+f_{22}(\vec{x}_i)$ in contrast to the heterogenous version (npivBART-h) introduced in Section~\ref{ssec:2sls+npivBART}. The priors for npivBART-g can be set in the same way as in the previous subsection, but applied to both $f_{21}$ and $f_{22}$.

Estimation for all of these models is trivial using a similar Bayesian backfitting approach outlined in the Supplementary Material for npivBART-h. However, the estimation of ivBART-h can be simplified by using a single regression tree ensemble for both $f_{21}(\vec{x}_i)$ and $f_{22}(\vec{x}_i)$ with a slight modification. Whereas usually a constant mean value is estimated in the terminal nodes $\M_h$, here a line with the exposure as the independent variable is estimated. The intercept and slope of this line correspond to $f_{21}(\vec{x}_i)$ and $f_{22}(\vec{x}_i)$. This method is, in principal, equivalent to semiparametric BART models that are particularly useful in a precision medicine context, as done in \cite{ZeldRe19}.

\subsection{Flexible Errors Using Dirichlet Process Mixture Models}
\label{ssec:dpm}

As noted in \cite{WiesHisg14} and ignoring the measured confounders $\vec{x}_i$ without loss of generality, $E[y_i|t_i,z_i]=f_2(t_i)+E[\epsilon_{y_i}|\epsilon_{t_i}]$. However, bivariate normality implies that $E[\epsilon_{y_i}|\epsilon_{t_i}]=\big([\Sigma]_{11}^{-1}[\Sigma]_{12}\big)\epsilon_{t_i}$ which is linear in $\epsilon_{t_i}$. Therefore, the consequence of a bivariate normality assumption is that the model assumes the unmeasured confounders are related to the outcome linearly. This is an unsatisfying assumption when the rest of the model is designed for flexible nonparametric estimation. As such, this assumption can be relaxed using a Dirichlet process mixture (DPM) model. The DPM model can be thought of as an infinite mixture of normals that can model \textit{any} probability distribution and so the induced linearity assumption is relaxed with such a specification. An additional benefit of this specification is that it is robust to outliers because it can model an error distribution with fatter tails than the usual normal errors.

This flexible error specification can be written as
\begin{equation}\label{eqn:dpm}
  \vec{\epsilon}_i\sim\sum_{i=1}^\infty\pi_{i}\text{N}_2(\vec{0}_2,\Sigma_{i})
\end{equation}
with $\Sigma_i\sim\mathcal{F}$ and $\mathcal{F}\sim\text{DP}(\mathcal{F}_0,\alpha)$ which defines a DPM model. $\mathcal{F}_0$ represents the base distribution for $\Sigma_i$ and $\alpha$ is the sparsity parameter. In this case, a two-dimensional inverse-Wishart distribution is used as $\mathcal{F}_0$. As detailed in the Supplementary Materials, sampling can be done using the algorithm of \cite{EscoWest95}.\nocite{Neal00}

\section{Simulation}
\label{sec:sim}

The simulation study aims to answer two separate issues concerning nonparametric instrumental variable regression using BART: (1) the overall effectiveness of using BART ensembles for an IV analysis and (2) the effect of using a Bayesian analysis for IV in the presence of weak instruments. Results from each simulation are presented in this section.

\begin{table}[!b]
  \centering
  \caption{The four true functions for $f_2$}
  \begin{tabular}{l|cc}
    & Global                    & Heterogenous            \\ \hhline{=|==}
    Linear    & $t_i+I(x_{1i}\ge 0)$       & $t_iI(x_{1i}\ge 0)$       \\
    Nonlinear & $\cos(t_i)+I(x_{1i}\ge 0)$ & $\cos(t_i)I(x_{1i}\ge 0)$\\
  \end{tabular}
  \label{tab:simf2}
\end{table}

To judge the effectiveness of the proposed method, data under eight varying simulation settings are generated $1{,}000$ times repeatedly. The sample size is set as $n=1{,}000$ and the variance of the errors is set as $V(\epsilon_{t_i})=V(\epsilon_{y_i})=1$. This implies that Cov$(\epsilon_{t_i},\epsilon_{y_i})=\rho$, the correlation. We set $\rho=0.0$ and $\rho=0.7$ to account for the situation where there are no unmeasured confounders and the situation where there are unmeasured confounders respectively. The regression function for the first stage is $f_1(\vec{z}_i)=\sin(\pi z_{1i}z_{2i})+2z_{3i}^2+z_{4i}+0.5z_{5i}$ is the traditional Friedman function \citep{Frie91}, commonly used for benchmarking nonlinear prediction problems. The results from transformation $f_1$ are scaled by their variance so that there is an equal amount of signal and noise in stage 1. Vector $\vec{z}_i$ was simulated according to the allele frequencies and correlation structure of the SNPs found in gene FTO, a gene associated with BMI and other cardiovascular related diseases. SNPs from this gene are used as the instruments for the causal analysis of UK Biobank data in Section~\ref{sec:ukbb}. 20 SNPs are simulated for $\vec{z}_i$. Ten additional independent covariates were simulated as $\vec{x}_i\sim\text{Unif}(-1,1)$. The functional form of the true second stage regression relationship is varied according to Table \ref{tab:simf2}.

\begin{figure}\centering
        \centering
        \includegraphics[width=\textwidth]{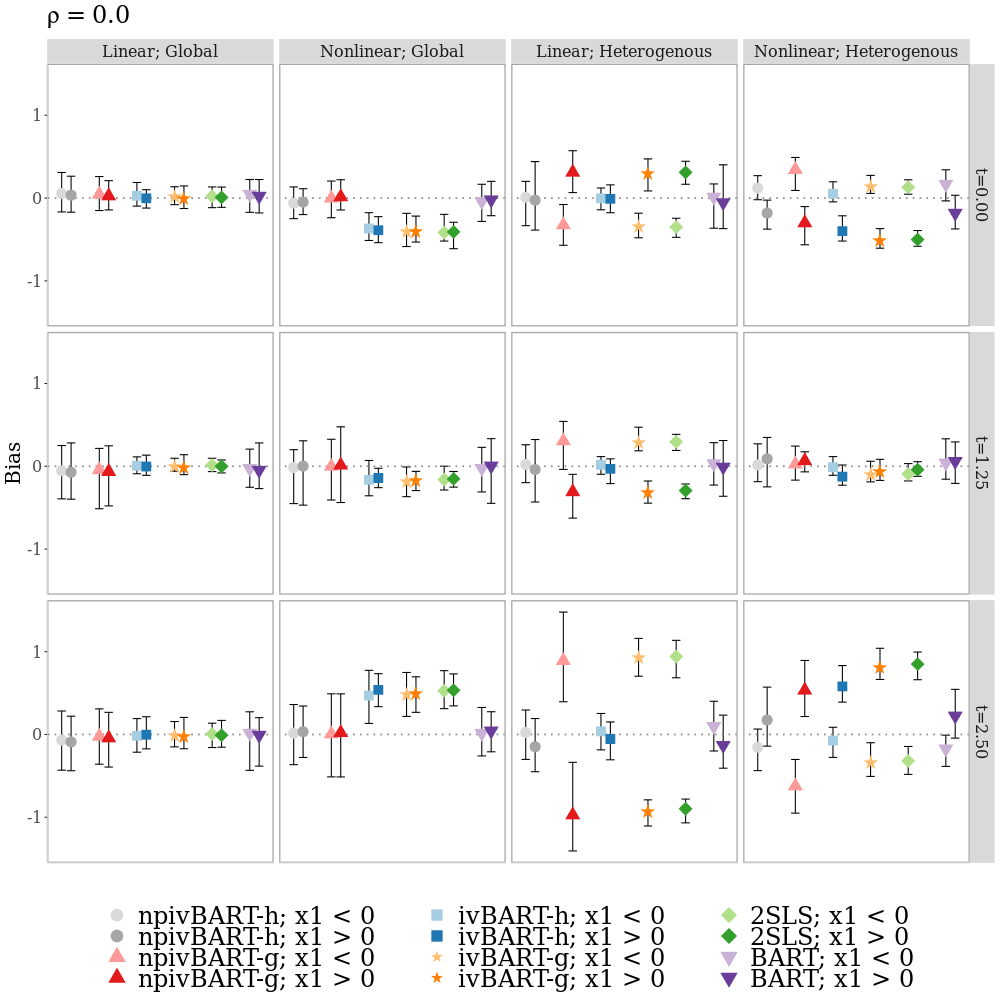}
        \caption{Simulation results showing the bias along the grid of $t\ge 0$ for the case without unmeasured confounding, i.e. $\rho=0.0$. Pictured are the mean and 95\% interval over $1{,}000$ simulated datasets. Traditional BART is unbiased because there is no unmeasured confounding. The most flexible npivBART-h is unbiased in all scenarios. Note the reduced variability in the linear methods when the underlying truth is linear.}
        \label{fig:rho0}
\end{figure}

Four versions of the method are considered, each corresponding to one of the true regression relationships above. These versions estimate: a global linear exposure effect (ivBART-g), a global nonlinear exposure effect (npivBART-g), a heterogenous linear exposure effect (ivBART-h), and a heterogenous nonlinear exposure effect (npivBART-h). The first three are the models discussed in Section~\ref{ssec:lessgeneral}. Additionally, two-stage least-squares (2SLS) and simple BART (fitting the second stage model without instruments) are also considered as comparators.

\begin{figure}\centering
        \centering
        \includegraphics[width=\textwidth]{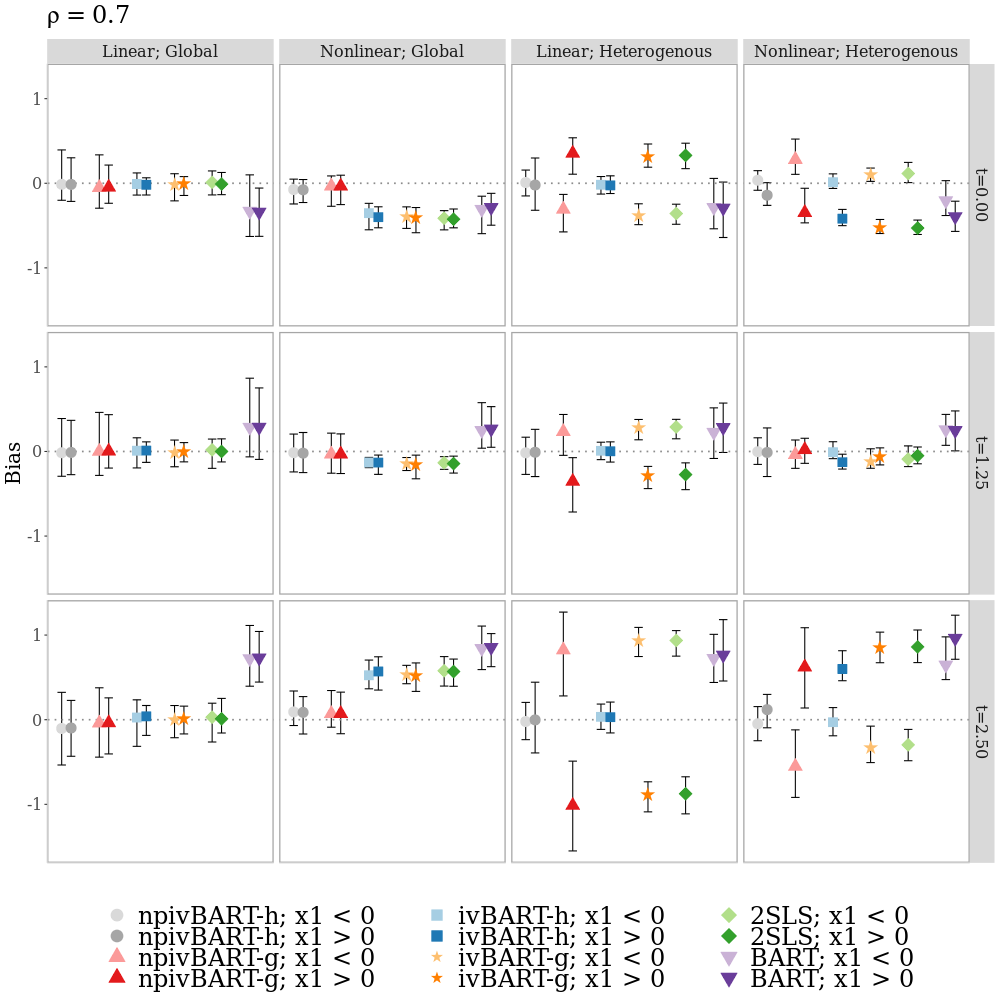}
        \caption{Simulation results showing the bias along the grid of $t\ge 0$ for the case with unmeasured confounding, i.e. $\rho=0.7$. Pictured are the mean and 95\% interval over $1{,}000$ simulated datasets. Traditional BART is biased when $t=1.25$ or $t=2.5$ implying that the confounding bias is worse in the tails where the data is sparse.}
        \label{fig:rho7}
      \end{figure}

For each model, $f_2(t_i,\vec{x}_i)$ is estimated for ten test points in the covariate space. These points are the five values of $t$ on an equidistant grid between $-2.5$ and $2.5$. These five points are then used in conjunction with $x_1=-0.5$ and $x_{1}=0.5$ (representing the $x_1<0$ and $x_1\ge 0$ cases respectively). The partial dependence function is used to margin out the $x_j$ for $j=2,\dots,p$. This yields ten estimates of $f_2$ which are compared to the true value of $f_2$ at each of these points. Then, the bias can be computed at each grid point while the RMSE can be computed across the five grid points for each of the positive and negative $x_1$. The bias results for $t\in\{0,1.25,2.50\}$ are given in Figure~\ref{fig:rho0} and Figure~\ref{fig:rho7} which give the results in the situation with $\rho=0.0$ and $\rho=0.7$. Table~\ref{tab:simres_err} gives the results for the RMSE across the grid points in every scenario. Note that the negative gridpoints $\{-1.25,-2.50\}$ show similar results, though the direction of the bias is usually flipped.

\begin{table}[!h]
  \caption{The mean (sd) RMSE for estimates of $f_2$ across the grid of $t$ values and $1{,}000$ simulations which mimic the bias results. However, it is of note that the linear methods perform slightly better than the nonlinear methods when the underlying truth is linear. This is because estimating linear coefficients is more stable than tree-based nonparametric estimators and this reduced variability yields reduced error (bias being equal).}\label{tab:simres_err}
          \begin{adjustbox}{max width=\linewidth}
            \begin{tabular}{lr@{\hspace{.5in}}rrrrrrrrrrrrr}\\
              \multicolumn{2}{l}{$\rho=0.0$}          & \multicolumn{1}{c}{npivBART-h} & \multicolumn{1}{c}{npivBART-g} & \multicolumn{1}{c}{ivBART-h} & \multicolumn{1}{c}{ivBART-g} & \multicolumn{1}{c}{2SLS} & \multicolumn{1}{c}{BART} \\\hline
              \multirow{2}{*}{Linear-g}    & $x_1=-0.5$ & 0.257 (0.05) & 0.193 (0.05) & 0.110 (0.05) & 0.085 (0.05) & 0.083 (0.05) & 0.221 (0.05) \\
                                           & $x_1=+0.5$ & 0.247 (0.06) & 0.195 (0.06) & 0.094 (0.06) & 0.087 (0.06) & 0.080 (0.06) & 0.211 (0.06) \\
              \multirow{2}{*}{Nonlinear-g} & $x_1=-0.5$ & 0.231 (0.05) & 0.227 (0.05) & 1.168 (0.05) & 1.127 (0.05) & 1.098 (0.05) & 0.218 (0.05) \\
                                           & $x_1=+0.5$ & 0.230 (0.04) & 0.222 (0.04) & 1.120 (0.04) & 1.130 (0.04) & 1.101 (0.04) & 0.212 (0.04) \\
              \multirow{2}{*}{Linear-h}    & $x_1=-0.5$ & 0.251 (0.08) & 1.059 (0.08) & 0.120 (0.08) & 0.961 (0.08) & 0.979 (0.08) & 0.296 (0.08) \\
                                           & $x_1=+0.5$ & 0.418 (0.12) & 0.917 (0.12) & 0.103 (0.12) & 0.924 (0.12) & 0.911 (0.12) & 0.390 (0.12) \\
              \multirow{2}{*}{Nonlinear-h} & $x_1=-0.5$ & 0.182 (0.03) & 0.415 (0.03) & 0.126 (0.03) & 0.369 (0.03) & 0.347 (0.03) & 0.222 (0.03) \\
                                           & $x_1=+0.5$ & 0.451 (0.08) & 0.465 (0.08) & 1.093 (0.08) & 0.924 (0.08) & 0.920 (0.08) & 0.442 (0.08) \\
               \multicolumn{2}{l}{$\rho=0.7$}           & \multicolumn{1}{c}{npivBART-h} & \multicolumn{1}{c}{npivBART-g} & \multicolumn{1}{c}{ivBART-h} & \multicolumn{1}{c}{ivBART-g} & \multicolumn{1}{c}{2SLS} & \multicolumn{1}{c}{BART} \\\hline
              \multirow{2}{*}{Linear-g}    & $x_1=-0.5$ & 0.207 (0.04) & 0.197 (0.04) & 0.099 (0.04) & 0.083 (0.04) & 0.093 (0.04) & 0.892 (0.04) \\ 
                                           & $x_1=+0.5$ & 0.199 (0.04) & 0.192 (0.04) & 0.080 (0.04) & 0.073 (0.04) & 0.087 (0.04) & 0.898 (0.04) \\ 
              \multirow{2}{*}{Nonlinear-g} & $x_1=-0.5$ & 0.174 (0.02) & 0.155 (0.02) & 1.159 (0.02) & 1.118 (0.02) & 1.079 (0.02) & 0.791 (0.02) \\ 
                                           & $x_1=+0.5$ & 0.165 (0.02) & 0.148 (0.02) & 1.096 (0.02) & 1.113 (0.02) & 1.074 (0.02) & 0.784 (0.02) \\ 
              \multirow{2}{*}{Linear-h}    & $x_1=-0.5$ & 0.223 (0.08) & 0.948 (0.08) & 0.083 (0.08) & 0.999 (0.08) & 0.984 (0.08) & 1.091 (0.08) \\ 
                                           & $x_1=+0.5$ & 0.259 (0.06) & 0.999 (0.06) & 0.095 (0.06) & 0.901 (0.06) & 0.917 (0.06) & 0.747 (0.06) \\ 
              \multirow{2}{*}{Nonlinear-h} & $x_1=-0.5$ & 0.194 (0.05) & 0.406 (0.05) & 0.081 (0.05) & 0.318 (0.05) & 0.316 (0.05) & 1.017 (0.05) \\ 
                                           & $x_1=+0.5$ & 0.315 (0.05) & 0.445 (0.05) & 1.068 (0.05) & 0.902 (0.05) & 0.897 (0.05) & 0.695 (0.05) \\ 

\end{tabular}
\end{adjustbox}
\end{table}

In particular, it is obvious that traditional BART, although it has as much flexibility as npivBART-h, cannot estimate $f_2$ without bias when $\rho=0.7$ as expected. The bias from BART is especially large for grid point $t=2.5$ in the sparse tail of the data. The model that is expected to correctly capture the regression relationship performs better than its less flexible counterparts. For example, ivBART-g and ivBART-h perform poorly when the relationship is nonlinear while ivBART-g and npivBART-g perform poorly when the relationship is heterogenous. 2SLS performs poorly with either nonlinearity or heterogeneity.

There does appear to be a slight benefit for choosing a linear model when the underlying truth is linear. Intuitively, a regression coefficient is more stable than estimating an ensemble of regression trees. More flexible nonlinear models still provide unbiased estimation in the linear case, but have an increased estimation variability. Table~\ref{tab:simres_err} presents the RMSE results with their standard errors and confirms that the linear methods have slightly reduced error than the nonlinear methods when the truth is linear, though both are unbiased.

An additional simulation was used to assess how weak instruments, one violation of the IV assumptions, may affect estimation. For simplicity, only the global linear variation of $f_2(t_i,\vec{x}_i)$ is considered as the true regression relationship. Likewise, only ivBART-g is considered so that the simulation criteria is the resulting estimated value of $\beta$. To mimic the weak instrument scenario, the true $f_1(\vec{z}_i)$ was scaled by a constant $C\in\{0.0,0.1,0.5,1.0\}$. A larger sample size of $n=5{,}000$ was considered in addition to the sample size of $n=1{,}000$ used in the first simulation.

The scalar $k$ represents the number of standard deviations covering the range of the data on the prior for the end predictions. This is described in Section~\ref{ssec:prior} as well as \cite{ChipGeor10}. For example, the default $k=2$ implies that central $95\%$ of the prior covers the range of the data while $k=3$ implies that the central $99.9\%$ of the prior covers the range of the data. Hence, increasing $k$ results in stronger regularization of the predictions (via stronger regularization of $\beta$) to $0$. The results in Table~\ref{tab:simres_beta} confirm that weak instruments will result in biased estimation. The BART-based methods slightly improve estimation of $\beta$ compared to 2SLS likely due to the regularization, but both are biased when $C$ is small. As expected, a large enough sample size could overcome this bias as indicated by the cases with small $C$ and $n=5{,}000$.

\begin{table}[!h]
  \centering
  \caption{The mean (sd) estimate of $\beta$ over $1{,}000$ replications from the weak instrument scenarios when $\beta=1$. Estimation appears to be biased in the weak instruments cases as expected, i.e. when $C=0.0$ or $C=0.1$. The value $k$ controls the amount of regularization through the amount of prior probability placed around $0$. A higher value of $k$ means that the resulting posterior distribution will be closer to $0$. Additionally, increasing the sample size results in less biased estimation as expected.}\label{tab:simres_beta}
  \begin{adjustbox}{max width=\linewidth}
  \begin{tabular}{lc@{\extracolsep{12pt}}cccccccc}
                          & \multicolumn{1}{c}{}              & \multicolumn{4}{c}{$n=1{,}000$}                                            & \multicolumn{4}{c}{$n=5{,}000$}                                        \\ \cline{3-6}\cline{7-10}
                          & \multicolumn{1}{c}{}              & $C=0.0$          & $C=0.1$         & $C=0.5$         & $C=1.0$          & $C=0.0$     & $C=0.1$     & $C=0.5$     & $C=1.0$ \\\cline{1-10}
    2SLS                  &  --                               & $1.67$ $(0.22)$  & $1.46$ $(0.17)$ & $1.06$ $(0.10)$ & $1.02$ $(0.04)$  & $1.73$ $(0.23)$ & $1.12$ $(0.09)$ & $1.01$ $(0.02)$ & $1.00$ $(0.01)$ \\
\multirow{1}{*}{ivBART-g} & $k=1$                             & $1.65$ $(0.15)$  & $1.56$ $(0.12)$ & $1.09$ $(0.09)$ & $1.04$ $(0.04)$  & $1.18$ $(0.26)$ & $1.14$ $(0.11)$ & $1.01$ $(0.02)$ & $1.00$ $(0.01)$ \\
                          & $k=2$                             & $1.57$ $(0.15)$  & $1.48$ $(0.13)$ & $1.07$ $(0.09)$ & $1.03$ $(0.04)$  & $1.18$ $(0.23)$ & $1.13$ $(0.11)$ & $1.01$ $(0.02)$ & $1.00$ $(0.01)$ \\
                          & $k=3$                             & $1.48$ $(0.16)$  & $1.39$ $(0.15)$ & $1.04$ $(0.10)$ & $1.01$ $(0.04)$  & $1.13$ $(0.24)$ & $1.11$ $(0.11)$ & $1.01$ $(0.02)$ & $1.00$ $(0.01)$ \\
                          & $k=4$                             & $1.37$ $(0.18)$  & $1.30$ $(0.16)$ & $1.02$ $(0.10)$ & $1.00$ $(0.04)$  & $1.05$ $(0.23)$ & $1.09$ $(0.11)$ & $1.00$ $(0.02)$ & $1.00$ $(0.01)$
  \end{tabular}
  \end{adjustbox}
\end{table}

\section{IV Analysis Using UKBB Data}
\label{sec:ukbb}

The large scale UK Biobank (UKBB) data which has approximately $500{,}000$ volunteers enrolled for study between the ages of 40 and 69 serves as a demonstration of the potential utility of this method. The centralized UK National Health Service allows for all health related events including disease diagnoses, drug prescriptions, and deaths to be recorded. Beginning in 2017, researchers were able to access the genetic information recorded with the database including genetic variants such as SNPs. As such, this database can be used to relate body mass index (BMI) as an exposure with systolic blood pressure (sBP) using single nucleotide polymorphisms (SNPs) as instruments.

BMI and blood pressure (systolic and diastolic) are traits known to have a strong relationship \citep{LandCalv18} that is thought to be nonlinear. Additionally, age and sex are two covariates that could be associated with either exposure or outcome. It would be interesting to explore whether the causal relationships differ based on either of these patient characteristics. These questions can be answered using the npivBART-h methodology introduced in this article.

To start, the summary statistics for systolic blood pressure (sBP) and diastolic blood pressure (dBP) outcomes by different groupings of BMI, age and sex are displayed in Table~\ref{tab:summ}. The summary statistics show a fairly large difference in the sBP distribution between the groups of BMI and age with a smaller difference between sexes. However, the summary statistics of dBP do not imply much of distributional difference in either age or sex. Therefore, dBP is not considered as an outcome for the rest of this section. Of course, these summary statistics do not speak to any potential interaction between the predictors, but that can be addressed when looking at the npivBART-h results.

 \begin{table}[!t]
    \centering
    \caption{Summary statistics for sBP and dBP by BMI, sex, and age. These summary results suggest an effect of all three on sBP. There does not appear to be a strong effect of either age or sex when looking at dBP as an outcome, however. As such, only sBP will be considered in the subsequent heterogenous analysis.}\label{tab:summ}\vspace{.1in}
    \begin{adjustbox}{width=\linewidth}
    \begin{tabular}{@{\extracolsep{16pt}}lrcccccccccc}
                                               &                      & \multicolumn{5}{c}{sBP}                                            & \multicolumn{5}{c}{dBP}                               \\\cline{3-7} \cline{8-12}
      Predictor                                & $n$                  & 1Q          &  Median    & Mean       & 3Q         & SD         & 1Q      & Median  & Mean      & 3Q      & SD          \\\hline
      \bf{Overall:}                            & $\bm{473{,}303}$     &  \bf{123} & \bf{135} & \bf{136.3} & \bf{148} & \bf{18.8}  & \bf{75} & \bf{82} & \bf{82.0} & \bf{89} & \bf{10.4}    \\
      \multicolumn{1}{l}{BMI:}                 &                      & \multicolumn{5}{c}{}                                            & \multicolumn{5}{c}{} \\
      \multicolumn{1}{l}{\hspace{.2in}$15-19$} & $4{,}449$            & 112        & 122        & 125.0       & 136        & 19.1       & 68      & 74      & 75.0      & 81      & 10.3 \\
      \multicolumn{1}{l}{\hspace{.2in}$20-24$} & $109{,}081$          & 117        & 129        & 130.8       & 142        & 18.9       & 71      & 77      & 78.0      & 84      & 10.0 \\
      \multicolumn{1}{l}{\hspace{.2in}$25-29$} & $213{,}690$          & 124        & 135        & 136.9       & 148        & 18.5       & 75      & 82      & 82.0      & 88      & 10.0 \\
      \multicolumn{1}{l}{\hspace{.2in}$30-34$} & $104{,}081$          & 127        & 138        & 139.8       & 151        & 18.1       & 78      & 84      & 84.7      & 91      & 10.0 \\
      \multicolumn{1}{l}{\hspace{.2in}$35-39$} & $30{,}416$           & 127        & 138        & 139.9       & 151        & 18.0       & 79      & 85      & 85.7      & 92      & 10.2 \\
      \multicolumn{1}{l}{\hspace{.2in}$40-44$} & $8{,}450$            & 127        & 139        & 140.2       & 151        & 18.2       & 80      & 86      & 86.5      & 93      & 10.4 \\
      \multicolumn{1}{l}{\hspace{.2in}$45+$}   & $3{,}136$            & 129        & 140        & 141.9       & 152        & 17.9       & 80      & 87      & 87.8      & 95      & 11.0 \\
      \multicolumn{1}{l}{Sex:}                 &                      & \multicolumn{5}{c}{}                                            & \multicolumn{5}{c}{} \\
      \multicolumn{1}{l}{\hspace{.2in}Female}  & $256{,}634$          & 120        & 132        & 133.7       & 146        & 19.3       & 73      & 80      & 80.5      & 81      & 10.2 \\
      \multicolumn{1}{l}{\hspace{.2in}Male}    & $216{,}669$          & 127        & 138        & 139.3       & 150        & 17.6       & 77      & 83      & 83.8      & 84      & 10.3 \\
      \multicolumn{1}{l}{Age:}                 &                      & \multicolumn{5}{c}{}                                            & \multicolumn{5}{c}{} \\
      \multicolumn{1}{l}{\hspace{.2in}$35-44$} & $48{,}621$           & 116        & 125        & 126.4       & 136        & 15.4       & 73      & 80      & 80.4      & 87      & 10.4\\
      \multicolumn{1}{l}{\hspace{.2in}$45-54$} & $133{,}910$          & 119        & 130        & 131.4       & 142        & 17.1       & 75      & 82      & 82.1      & 89      & 10.5\\
      \multicolumn{1}{l}{\hspace{.2in}$55-64$} & $200{,}193$          & 126        & 137        & 138.6       & 150        & 18.6       & 75      & 82      & 82.4      & 89      & 10.3\\
      \multicolumn{1}{l}{\hspace{.2in}$65+$}   & $90{,}579$           & 130        & 142        & 143.7       & 155        & 19.3       & 75      & 81      & 81.5      & 88      & 10.3\\
    \end{tabular}
    \end{adjustbox}
  \end{table}

  Gene FTO on chromosome 16 is known to be associated with obesity related disorders and contains the first genetic variant to be associated with BMI. Discovered in 2007, subsequent studies have discovered additional genetic variants from FTO that are also strongly associated with BMI and this effect appears to be consistent among different ethnic populations. SNPs from the cis-region of this gene, defined as 50kbp on each side of the gene, of FTO were extracted as candidate instruments for this analysis. Of these extracted genes, $p_Z=8$ SNPs were used as instruments because they have been identified as risk variants for BMI in a genome-wide replication study \citep{TanZhu14}. Such genetic variants should not be associated with environmental confounding factors, motivating their use as instruments and is commonly called Mendelian randomization in the statistical genetics literature.
  
\begin{figure}[!t]
  \centering
  \includegraphics[width=\textwidth]{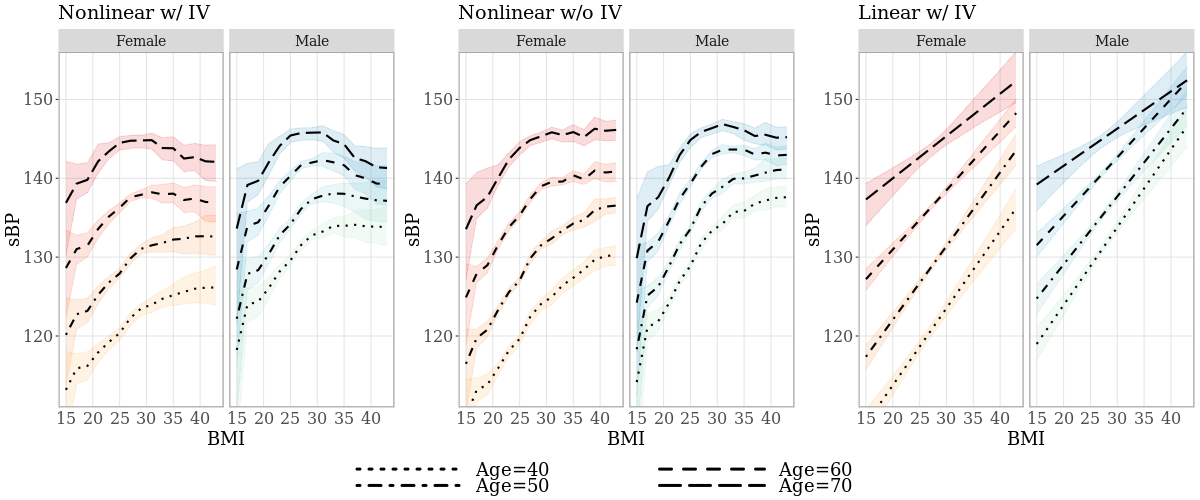}
  \caption{The estimated curves for the nonlinear IV (left), nonlinear non-IV (middle), and linear IV (right) methods. The nonlinear estimates give a more complete picture of the relationship, particularly near BMI values in the tails. The bands represent the 95\% credibility interval for the estimated $f_2$.}\label{fig:UKBB_age}
\end{figure}

In total, $473{,}303$ observations were used from the UKBB dataset. The Dirichlet Process Mixture (DPM), defined in Equation~\eqref{eqn:dpm}, was used as a flexible error distribution. Three models are used for comparison in this section. The first is an IV regression that is nonlinear in BMI using npivBART-h, the second is a nonlinear regression without IVs using traditional BART, and the third is an IV regression that is linear in BMI using ivBART-h. For all methods, there were $5{,}000$ burn-in samples and $5{,}000$ additional samples drawn from the posterior distribution across three chains leading to a total of $15{,}000$ posterior samples. Each chain utilized different starting values to promote proper mixing. Thinning the MCMC sample by a factor of $k$ to reduce the autocorrelation, while recommended in the past, is now recognized as not being beneficial because reducing the total number of posterior draws by a factor of $k$ is more detrimental than autocorrelation in most cases \citep{LinkEato12}. As such, no thinning is performed for these analyses.

The resulting curves given in Figure~\ref{fig:UKBB_age} were plotted for a grid of fifteen points of BMI, four ages, and both sexes. The nonlinear IV method (left) utilizes a nonlinear second stage for the IV regression. First, it appears that there may be a three-way interaction between BMI, age, and sex. For the male observations in particular, the effect of age is much larger for moderate values of BMI compared to large values of BMI where the age effect is reduced. This two-way interaction does not appear in the female group leading to the three-way interaction. When comparing the nonlinear IV method to its non-IV counterpart (middle), there appears to be a difference in the estimated curves, particularly in the tail values of BMI. The non-IV method tends to estimate a steeper increasing trend overall. This seems reasonable because the non-IV method is confounded by environmental factors, making the effect of BMI appear stronger than it truly is. Finally, the linear IV method (right) is considered as comparison to the nonlinear methods. This plot suggests a much stronger three-way interaction than its nonlinear comparators. The rigidity of linear estimates are apparent here because the predictions for higher BMI values are overwhelmed by the observations with moderate BMI values where the increasing effect is strongest. In cases such as this, nonlinear regression methods will paint a more complete and nuanced picture of the true relationship.

\begin{figure}[!b]
  \includegraphics[width=\textwidth]{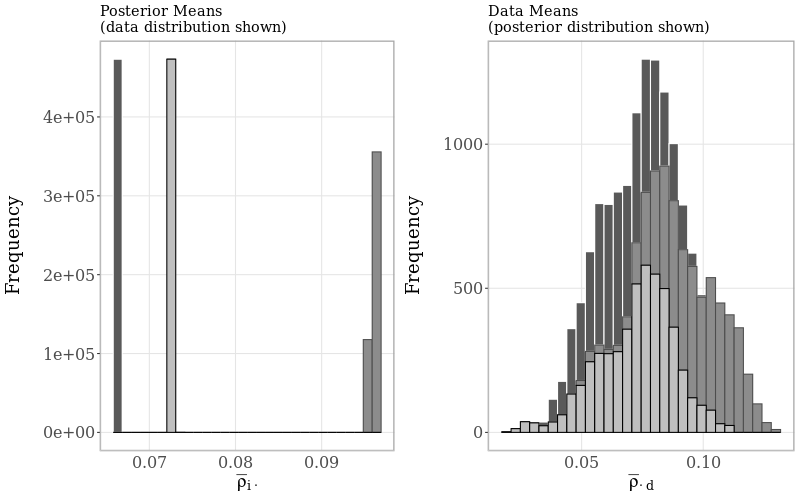}
  \caption{The distribution of posterior means across the data (left) and the distribution of data means across the posterior (right) for the error correlation $\rho$. The three chains are indicated with shading. The left plot shows little heterogeneity within the error correlation among the observations. As such, averaging over the sample as displayed in the right yields a posterior distribution that indicates $\rho>0$.}\label{fig:cor}
\end{figure}

While a difference between the IV and non-IV methods appears visually, it can be confirmed by looking at the estimated error correlation. The relationship is confounded by unmeasured variables when this correlation is non-zero. However, because the DPM is used as a prior for the error distribution, each observation has its own $2\times 2$ variance-covariance matrix estimated at each draw of the MCMC chain. Let the correlation term from this variance-covariance matrix be denoted as $\rho_{id}$ for observation $i$ at MCMC iteration $d$. Let averages of these estimates be defined as $\mean{\rho}_{\cdot d}=(1/n)\sum_{i=1}^n \rho_{id}$ and $\mean{\rho}_{i\cdot}=(1/D)\sum_{d=1}^D\rho_{id}$ where $D=5{,}000$ in this case. Note that $\mean{\rho}_{i \cdot}$ are computed within each chain for visualization. Histograms of these distributions are given in Figure~\ref{fig:cor} with each of the three chains distinguished using the shading. The left panel shows little correlation heterogeneity among the observations within each chain, suggesting that a global estimate of this correlation is sufficient. In fact, most iterations of the MCMC chain resulted in less than 3 clusters being detected among the observations. Because of this, averaging over the observations to yield the posterior distribution of a single correlation term will be valid. The right panel of the figure, which shows this posterior distribution, indicates that the error correlation is positive and non-zero, though relatively small. This is in line with the relatively small visual difference seen between the IV and non-IV curves of Figure~\ref{fig:UKBB_age}.

\section{Discussion}
\label{sec:conc}

This article introduces npivBART-h, a highly flexible instrumental variable framework using Bayesian additive regression trees capable of estimating the individualized nonlinear relationship between an exposure and outcome in the presence of unmeasured confounding. Simulation results confirm that this is a feasible method for estimating heterogenous nonlinear regression relationships with a causal interpretation. With the increasingly large volume of complex observational data being gathered today, statistical methods to flexibly estimate the regression relationship while also accounting for unmeasured confounding will necessary. These nonlinear estimates can give researchers a more thorough set of results. Indeed, there are a variety of applications where applying flexible methods such as npivBART-h could prove scientifically interesting. This article presents one such application in the area of statistical genetics and Mendelian randomization where genetic variants are used as instruments to compute a causal effect of traits on an outcome. As done in Section~\ref{sec:ukbb}, body mass index can be causally related to blood pressure, a relationship that is commonly thought to be nonlinear.

Additionally, the method allows for individualized heterogenous estimation which could prove useful for tailoring treatments to the characteristics of the patient, usually known as precision medicine. This is an exciting new frontier in the medical field and could lead to individualized treatment strategies that improve health outcomes for society at large. Developing analysis tools such as npivBART-h to execute this type of research is an opportunity for statisticians to meaningfully contribute to this potential paradigm shift.

As mentioned in Section~\ref{sec:intro}, the Bayesian probability framework has many benefits. First, it allows for easy extendibility through hierarchical models so that more complex data can be analyzed. One simple extension to our method would be to allow more complicated outcomes such as binary, count, or time to event either for the exposure or outcome. These types of outcomes have already been considered for traditional BART and so this extension would be a relatively straightforward way to broaden the method's applicability. Additionally, sparsity based methods such as \cite{RockVand17} and \cite{Line18} should be considered to exclude potential weak instruments in the first stage which can bias the results as shown through simulation. While it may boost the predictive power of the model, including additional instruments increases the likelihood of IV assumption violations. It is for this reason that a parsimonious set of SNPs is to be preferred, something that can be achieved using sparsity methods.

One interesting application in statistical genetics would be to apply this methodology across an entire genome, relating the expression levels of a gene to an outcome in a possibly nonlinear fashion. In this case, the expression levels of each gene use the SNPs within their cis-region as instruments. Doing this across the entire genome is known as a transcriptome-wide association study (TWAS) and is usually based on linear 2SLS methods. However, there is precedent for using machine learning prediction models for this type of analysis \citep{OkorSchu21}. For an overview of TWAS and additional statistical considerations, see \cite{XuePan20} or \cite{LiRitc21}.

An R software package entitled \texttt{hnpivbart} is available to fit any of these models. This implementation of the BART algorithm is based off of the \texttt{BART3} package \citep{SparSpan21} which utilizes C++ code for fast and efficient operation of the method.

\section*{Acknowledgements}
The authors would first like to acknowledge the Minnesota Supercomputing Institute at the University of Minnesota for providing resources that contributed to the research results reported within this paper. \url{http://www.msi.umn.edu}.

\section*{Funding}
This work was supported by the National Institutes of Health grant R01HL116720.

\bibliographystyle{apa-good-unsrt} 
\bibliography{refs}       

\end{document}